\documentclass{llncs}
\usepackage{graphicx}

\newcommand{\statementid}[1]{\textsf{\footnotesize#1}}

\begin{document}

\title{
  How to Evaluate Controlled Natural Languages
}

\author{
  Tobias Kuhn
}

\institute{
  Department of Informatics, University of Zurich, Switzerland\\
  \texttt{tkuhn@ifi.uzh.ch}\\
  \texttt{http://www.ifi.uzh.ch/cl/tkuhn}
}

\maketitle

\begin{abstract}
This paper presents a general framework how controlled natural languages can be evaluated and compared on the basis of user experiments. The subjects are asked to classify given statements (in the language to be tested) as either true or false with respect to a certain situation that is shown in a graphical notation called ``ontographs''. A first experiment has been conducted that applies this framework to the language Attempto Controlled English (ACE).
\end{abstract}

\section{Introduction}

Controlled natural languages (CNL) are claimed to be easier to learn and understand than other formal languages. User studies are the only way to verify this claim. Many such studies have been conducted on tools that use some kind of CNL, e.g. the one described in \cite{bernstein06gino}. However, it is hard to determine in these cases how much the CNL contributes to the understandability and how much is due to other aspects of the tool. Furthermore, it is hard to compare CNLs to other formal languages with such studies because different languages often require different tools. For those reasons, it would be very desirable to be able to test a CNL in a tool-independent way.

\cite{hart:eswc2008} presents one of the rare attempts to test a CNL (concretely the Rabbit language that is based on English) independently of a particular tool. The authors conducted an experiment where the subjects were given one Rabbit statement at a time and they had to choose from four paraphrases in natural English only one of which was correct. There are two problems with this approach. First, since natural language is highly ambiguous, we have to make sure somehow that the subjects understand the natural language paraphrases in the right way, which just takes the same problem to the next level. Second, since the formal statement and the paraphrases look very similar in many cases (both rely on English), it is hard to determine whether understanding is actually necessary to fulfill the task. The subjects might do the right thing without understanding the sentences (e.g. just by following some syntactic patterns), or by misunderstanding both --- statement and paraphrase --- in the same way.

\section{Approach}

\begin{figure}[p]
  \begin{center}
    \includegraphics[width=\textwidth]{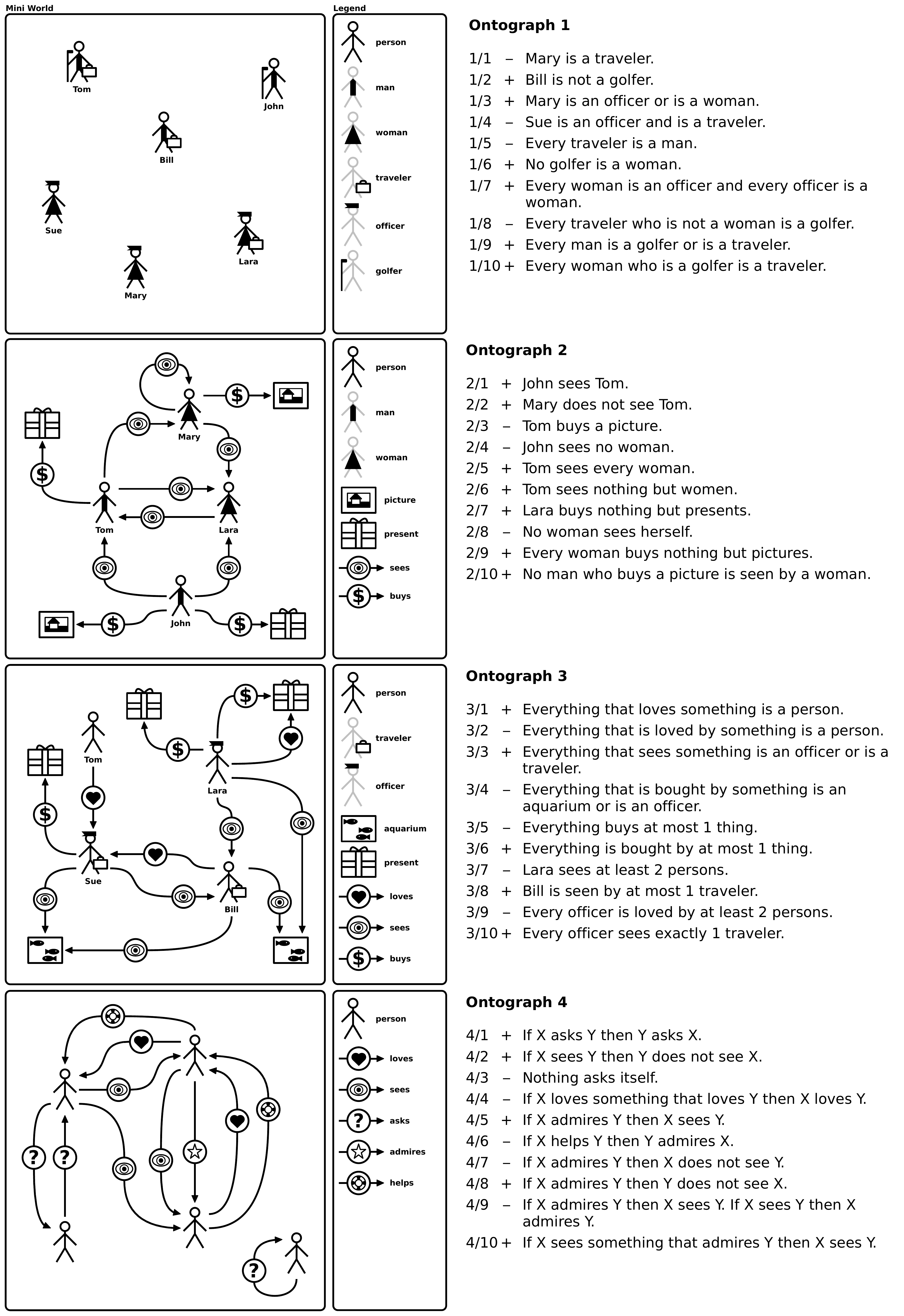}
    \caption{The four ontographs that have been used for the experiment together with the respective ACE statements. True statements are marked by +, false ones by --.}
    \label{fig:ontographs}
  \end{center}
\end{figure}

The approach that we propose here solves the discussed problems and relies on a graphical notation which we call ``ontographs''.  Each ontograph consists of a legend that introduces types and relations and of a mini world that introduces individuals, their types, and their relations. Individuals are represented by different symbols depending on the types they belong to. Relations are represented by arrows that point from one individual to another. Figure \ref{fig:ontographs} shows four such ontographs. Ontographs require that we have complete information about the mini world and, as a consequence, do not need an explicit notation for negation. Everything that is not shown in the ontograph is not true.

We assumed that ontographs are very easy to understand and our results confirm this assumption. The use of intuitive graphical icons is one of the reasons of the good understandability. More important, however, is the fact that the ontograph notation has no generalization capabilities and does not support partial knowledge. This excludes many potential misunderstandings. An ontograph explicitly shows every existing individual and depicts every single relation instance between them. There is not much to misunderstand with these basic elements. While we can make general statements like ``every man loves a woman'' about the mini world defined by an ontograph, there is no way to express such statements in this general way in the ontograph notation itself.

Ontographs are designed to be used in experiments to test the understandability of formal languages. Ontographs could also be used to test the writability by asking the subjects to describe the given situation. However, only the first approach has been investigated so far.

In order to test the understandability of a language in an experiment, an ontograph and several statements (written in the language to be tested) are shown to the subjects who have to decide which of the statements are true and which are false with respect to the mini world depicted by the ontograph.

An important property of ontographs is that they use a graphical notation that is syntactically very different from textual languages like CNLs. This makes it virtually impossible to accomplish a task like the one described above just by looking at the syntax. If subjects manage to systematically classify given statements correctly as true or false with respect to a certain ontograph then we can conclude that the subjects understood the statements and the ontograph.

Using the presented testing framework, we conducted a small experiment that tests the language Attempto Controlled English (ACE) \cite{short_fuchs:reasoningweb2008} which is a mature controlled natural language. The goal was to find out how well the framework works and at the same time how understandable ACE is. We recruited 15 subjects who were not experts in knowledge representation.

Four ontographs (see Figure \ref{fig:ontographs}) --- each accompanied by 10 ACE statements --- were shown to the subjects who had to classify each of the statements as true, false, or ``don't know''. There was a time limit of five minutes for each ontograph. There was no explanation how the ACE statements have to be interpreted with the one exception that it was explained that ``something'' can stand for persons and objects.

The ACE sentences are chosen in a way that they cover a broad variety of semantic structures. All sentences would be expressible in OWL and would cover most of the axiom types provided by the OWL standard. The first ontograph contains only individuals and types but no relations. The second ontograph introduces relations. The statements of the third ontograph use more complicated structures like domain, range, and cardinality restrictions. The statements of the fourth ontograph, finally, use no individuals and no types but talk only about relations. In this way, we cover a subset of ACE that corresponds to a subset of first-order logic that is similar to the one used by OWL.

\section{Results}

\begin{figure}[tb]
  \begin{center}
    \includegraphics[width=\textwidth]{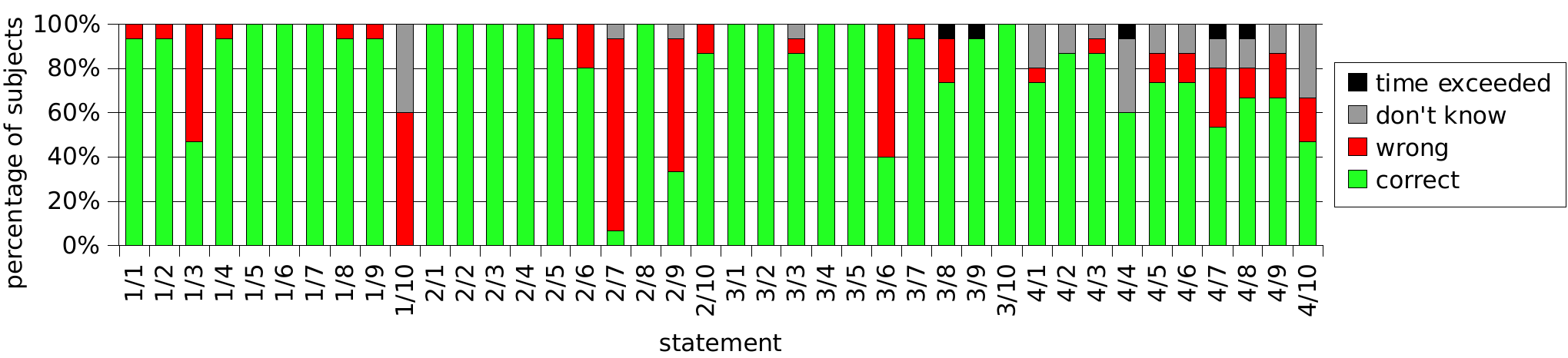}
    \caption{This chart shows for each statement how many subjects classified it correctly, classified it wrongly, said that they don't know, or exceeded the time limit.}
    \label{fig:table}
  \end{center}
\end{figure}

Figure \ref{fig:table} shows the result of the experiment. Overall, the decision rate (i.e. the percentage of subjects who classified a particular sentence as either true or false) was 93\%, and on average each decision took 21 seconds. The decisions were correct in 85\% of the cases (compared to 50\% which can be achieved by mere guessing). The chart shows that the incorrect decisions are not equally distributed. In the case of five statements the incorrect behavior is predominant. The statements \statementid{1/3}, \statementid{1/10}, \statementid{2/7}, and \statementid{2/9} represent special cases and for this reason it is not unexpected that they are often classified wrongly.

The statement \statementid{1/3} is a very simple statement using ``or'' that is true but was often classified as false. One could think that many subjects misinterpreted the ``or'' as being exclusive (instead of inclusive as ACE defines it). However, if that is the case then the subjects should also interpret the ``or'' of statement \statementid{1/9} as exclusive, but they did not. A more plausible explanation is that the subjects recognized that the statement \statementid{1/3} is imprecise in the sense that using ``and'' instead of ``or'' would be more accurate. This prevented some of the subjects from realizing that the statement is nevertheless true in a logical sense. Things are different with statement \statementid{1/9} where the replacement of ``or'' by ``and'' would not be more accurate but would make the true statement false. As a result, \statementid{1/9} was classified correctly by almost all subjects. It seems that people in such cases often fail to distinguish accuracy from logical truth.

The statement \statementid{1/10} is another special case. It is a conditional statement with a false precondition. It is not surprising that people with no background in logics fail to classify this statement in a correct way.

\statementid{2/7} and \statementid{2/9}, finally, are common mistakes that were also encountered in OWL: people tend to think that ``nothing but presents'' implies ``at least 1 present'' \cite{short_rector:ekaw2004}.

We assume that these types of mistakes are not specific to ACE, but can be encountered in any other language. Disregarding those four special cases, we get a decision rate of 94\% with 92\% correct decisions.

Since all three sentences containing ``nothing but'' were often misunderstood, we will investigate how this can be improved. Also the statement \statementid{3/6} requires more investigation.

Apart from the five exceptional statements, the correct decision was chosen at least twice as often than the wrong one. This preference is statistically significant on a 5\% level for all statements except \statementid{4/7} and \statementid{4/10} (the null hypothesis being a random 50\% decision).

Another interesting point is that we did not tell the subjects explicitly that the ontograph notation shows the complete information about the mini world and that everything that is not shown in the ontograph can be considered false. The result for statement \statementid{2/2} shows that the subjects understood this very well without any explanation. They understood that the fact that the ontograph does not show a sees-relation from Mary to Tom means that ``Mary does not see Tom'' is a true statement.


\section{Conclusions}

We presented a framework to test controlled natural languages relying on the graphical notation of ontographs. We applied this framework to test ACE and come to the conclusion that most ACE sentences are understood very well and very quickly. These results and individual discussions with the subjects let us conclude that ontographs are a suitable and powerful concept for understandability evaluations of CNLs. A larger and more thorough experiment that compares ACE to another formal language is currently being set up.

\bibliography{attempto}
\bibliographystyle{plain}

\end{document}